# Stream Processor Generator for HPC to Embedded Applications on FPGA-based System Platform

Kentaro Sano, Hayato Suzuki, Ryo Ito, Tomohiro Ueno, and Satoru Yamamoto
Graduate School of Information Sciences,Tohoku University
6-6-01 Aramaki Aza Aoba, Sendai 980-8579, JAPAN
Email: kentah@caero.mech.tohoku.ac.jp

*Abstract*—**This paper presents a stream processor generator, called SPGen, for FPGA-based system-on-chip platforms. In our research project, we use an FPGA as a common platform for applications ranging from HPC to embedded/robotics computing. Pipelining in application-specific stream processors brings FPGAs power-efficient and high-performance computing. However, poor productivity in developing custom pipelines prevents the reconfigurable platform from being widely and easily used. SPGen aims at assisting developers to design and implement high-throughput stream processors by generating their HDL codes with our domain-specific high-level stream processing description, called SPD. With an example of fluid dynamics computation, we validate SPD for describing a real application and verify SPGen for synthesis with a pipelined data-flow graph. We also demonstrate that SPGen allows us to easily explore a design space for finding better implementation than a hand-designed one.**

## I. INTRODUCTION

Advancement of FPGA technologies has made custom computing machines more promising and feasible for a wide range of applications from high-performance computing (HPC) to embedded computing. Dedicated pipelines implemented on reconfigurable devices can achieve higher performance per power with efficient use of a limited bandwidth for external memories. In addition to the FPGA's prominent performance of non-arithmetic applications, e.g., pattern matching with regular expressions, state-of-the-art FPGAs overcome general-purpose microprocessors even for floating-point computations. It is announced that the next generation of FPGAs will have 10 GFlops [1], which is higher than that of a present GPU. Thus FPGAs are expected to bring a breakthrough even in supercomputing where conventional approaches with CPUs and GPUs are suffering from a low efficiency of bandwidth and power [14]. Furthermore, the tightly-coupled architecture with FPGA fabric and a hard-core ARM processor [2] also allows more flexible and all-in-one computation especially fit for embedded/robotics applications.

In our research project, we have developed an FPGA-based common system-on-chip (SoC) platform for power-efficient computation in the wide range of applications. The platform has a microprocessor, peripherals including a PCI-Express interface and network interfaces with remote DMA engines, and reconfigurable stream processors with their local DMA engines. The stream processors are application-specific computing pipelines to accelerate kernels of a target application. The microprocessor controls the DMA engines and the peripherals while it can also perform irregular and auxiliary computations. If the performance has to be scaled with multiple FPGAs, we can give the SoC platforms a dedicated network to directly connect FPGAs.

So far, we have demonstrated feasibility of the FPGA-based power-efficient computation with the SoC platform and stream processors for floating-point computations [8], [11], [12]. However poor productivity is still a big issue in developing application-specific pipelines. Implementation with a hardware description language (HDL) is not easy, requiring a sufficient skill and a long time. For example, development of the stream processor for fluid dynamics simulation [8], [12] required designing and implementing sub-modules each of which is for a part of an entire stream computation, a top module to connect the sub-modules with adequate delay insertion, and interface logics for the SoC platform. These steps are not only tedious and time-consuming, but also error-prone. To concentrate on algorithm design, we need a tool to generate an application-specific stream processor with high-level description.

In this paper, we present a stream processor generator, SPGen, which generates a stream processor for the SoC platform with high-level stream processing description, called SPD. In SPD, we simply write numerical equations and calls of user-defined hardware modules. SPGen gives the processor a stream interface to be used in embedding it into the SoC platform. We can specify a target frequency to change the pipeline depth for design space exploration under a trade-off between area and throughput. Contributions of this work are:

1) SPGen to generate a stream processor for floating-point computations with equations and user-defined modules,
2) domain-specific stream processing description (SPD),
3) verification and evaluation with practical computation.

This paper is organized as follows. Section II summarizes related work. Section III describes the FPGA-based SoC platform, a stream computing model, and a stream processor. Section IV presents SPGen after introducing the SPD format. Section V shows verification and evaluation of SPGen with an example of fluid computation based on the lattice Boltzmann method. Finally, Section VI gives conclusions and future work.

## II. RELATED WORK

Recent advancement has allowed FPGAs to be programmed with high-level abstraction, by supporting conventional programming languages including C and Java with some extension or limitation. Tools or compilers such as Impulse-C [7], AutoPilot [15], Synphony [13], and a formal design framework [6] can be used to implement an IP core in high-level languages instead of describing RTL in HDL. Although their abstraction of description is useful, SPGen is more specific to a domain of stream computation with floating-point operators to easily develop high-throughput pipelines of numerical applications. We also leave the door open to extend a stream processor beyond just a computing pipeline by including user-defined modules. Compilers such as FCUDA [9], MaxCompiler [10], and OpenCL for FPGA [1] can compile a software program into high-performance floating-point pipelines which operate on an FPGA. However they are designed for FPGA-based accelerators, like GPUs, attached to a host CPU. In this sense, they are not suitable for IP core generation while our SPGen is designed to easily generate IP cores of stream processors that can be embedded into a common SoC platform on an FPGA. Furthermore, based on the first version of SPGen, we are going to develop a compiler for structured hardware such as a systolic array, which is more than just streaming pipelines, with its abstract description.

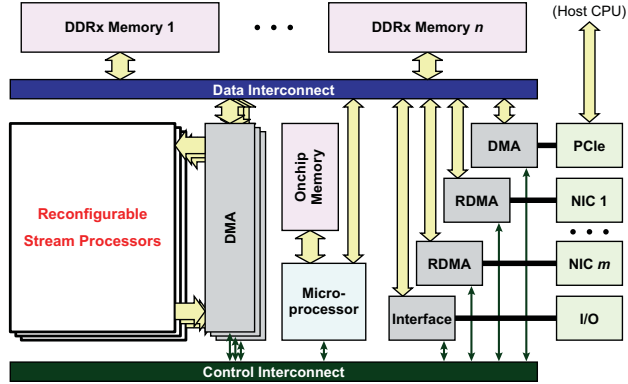

Fig. 1. FPGA-based common system-on-chip platform.

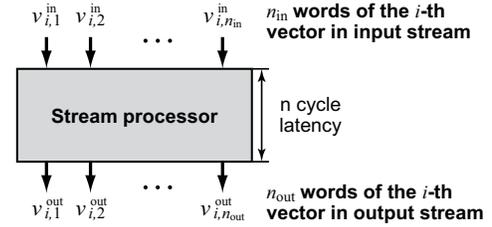

Fig. 2. Definition of a stream processor.

## III. FPGA-BASED COMMON SYSTEM PLATFORM AND STREAM PROCESSOR

### A. FPGA-based common system-on-chip platform

In our research project, we use FPGAs as power-efficient accelerators for a wide range of applications from HPC to embedded computing on robots and unmanned vehicles. For this purpose, we have been developing an FPGA-based common SoC platform shown in Fig.1, by using the system integration tool "Qsys" by ALTERA corporation. Qsys allows us to implement an on-FPGA system by simply connecting IP cores with memory-mapped or stream interfaces. The platform has reconfigurable stream processors, a microprocessor, peripherals, and external DDR memories, which are connected with the data and control interconnects.

The reconfigurable stream processors accelerate major and regular computation of target applications. Application-specific computing pipelines are statically, or dynamically if partial reconfiguration is available, configured to provide high-throughput stream computation. The stream data are supplied by the direct-memory access (DMA) modules connected to the external DDR memories via the data interconnection. We have to design custom computing hardware of stream processors for an individual application.

The microprocessor executes software to control the DMA and the peripherals in addition to performing irregular computation with data stored in the external memories. For conventional FPGAs, we implement a soft-core processor while such a hard-core processor as ARM Cortex A9 processor is available on a recent FPGA SoC [2]. We have several peripherals whose necessity depends on system purposes, including a PCI-Express (PCIe) interface with DMA, network interfaces (NICs) such as 10G Ethernet with remote DMAs, and I/O modules. The PCIe is necessary if the FPGA operates as an slave of a host PC. The NICs are used to scale the system and the performance over multiple FPGAs directly connected by their dedicated network. The I/O modules are necessary for robotics applications receiving data from sensors and controlling actuators. Reconfigurability of FPGAs allows us to implement only necessary modules on the system under resource constraint.

### B. Stream computing model and stream processor

We define data streams $S^{\text{in}}$ and $S^{\text{out}}$ as follows, which correspond to inputs and outputs of computation, respectively.

$$S^{\text{in}} \equiv \{v_0^{\text{in}}, v_1^{\text{in}}, ..., v_i^{\text{in}}, ..., v_n^{\text{in}}\} \quad (1)$$
$$S^{\text{out}} \equiv \{v_0^{\text{out}}, v_1^{\text{out}}, ..., v_i^{\text{out}}, ..., v_n^{\text{out}}\}, \quad (2)$$

where each element of these series is a vector: $v_i^{\text{in}} \equiv \{v_{i,1}^{\text{in}}, v_{i,1}^{\text{in}}, ..., v_{i,n_{\text{in}}}^{\text{in}}\}$ and $v_i^{\text{out}} \equiv \{v_{i,1}^{\text{out}}, v_{i,1}^{\text{out}}, ..., v_{i,n_{\text{out}}}^{\text{out}}\}$. With the input and output data-streams, the stream computation that we use is modeled by $v_i^{\text{out}} = f\left(v_{i-M}^{\text{in}}, ..., v_i^{\text{in}}, ..., v_{i+N}^{\text{in}}\right)$, which means that the $i$-th output is obtained by computing with $(N+M+1)$ input vectors around the $i$-th input. If $v_i^{\text{out}}$ can be computed only with $v_i^{\text{in}}$, then $M = N = 0$. The function $f()$ can contain any numerical computations or other processing. A simple example with an input vector $v_i^{\text{in}} = \{a, b, c, d\}$ and an output vector $v_i^{\text{out}} = \{lg, sm\}$ is:

$$tmp1 = (a-b)/2, \quad (3)$$
$$tmp2 = tmp1/c + d, \quad (4)$$
$$(lg, sm) = \begin{cases} (tmp2, tmp1) & (tmp1 < tmp2) \\ (tmp1, tmp2). & (\text{otherwise}), \end{cases} \quad (5)$$

where $tmp1$ and $tmp2$ are temporal variables used internally.

Fig.2 shows a stream processor to compute $v_i^{\text{out}}$ with $v_i^{\text{in}}$. The inputs of $v_k^{\text{in}}$ for $k \neq i$ are obtained by using an internal buffer to hold input vectors for a certain number of cycles. The stream processor is a computing pipeline with a throughput of one. Vectors are input and output to/from the stream processor synchronously one by one every cycle. The computation takes $n$ cycles for $n$ pipeline stages. The more stages we have, the more computations can be performed at a constant throughput. Thus the higher computing performance can be achieved by unrolling a computing kernel to increase the number of operations in a pipeline.

The stream processor has an interface to be embedded into the SoC platform of Fig.1. We adopt an interconnection standard, called Avalon-ST, provided in Qsys system integration tool. Avalon-ST is simple and light-weight interface for data streaming, which includes data and control signals of valid, ready, start of packet (sop), end of packet (eop), and empty bytes in data. For Avalon-ST interface, the stream processor

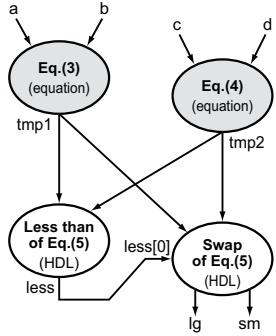
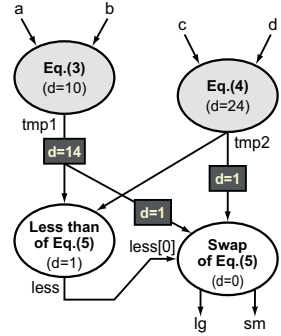

Fig. 3. Connection of modules before delay insersion.

Fig. 4. Connection of modules after delay insertion.

```
 1 : #-----------------------------------------
 2 : #---- Sample spd code --------------------
 3 : Name    sample_core
 4 : Input   a, b, c, d;
 5 : Output  lg, sm;
 6 : Param   p1 = 0.500;   # 1/2
 7 :
 8 : #---- equation nodes --------------------
 9 : eq1     0, equ,  tmp1    = (a - b) * p1;
10 : eq2     0, equ,  tmp2    = temp1 / c + d;
11 :
12 : #---- HDL nodes --------------------
13 : lsthan  1, HDL,  (less)  = less_than(tmp1, tmp2);
14 : swap    0, HDL,  (lg, sm) = swap(less[0], tmp1, tmp2);
```

Fig. 5. Stream-processing description (spd) code for equations (3) to (9). A line beginning with "#" is a comment. "Param" defines constant FP values.

is composed of the stream processing core and the other peripheral modules to handle the signals.

### C. Conventional development flow of stream processors

So far, the following development flow has been required:

1) design a stream algorithm as a data flow graph (DFG)
2) design and implement modules of nodes in DFG,
3) connect modules with delays inserted for synchronization,
4) write a top module with Avalon-ST interface,
5) write a configuration file in tcl for Qsys.

Let's suppose a simple stream processor of Fig.3 for Eqs.(3) to (5). In this example, the fundamental modules in DFG are the nodes: Eq.(3), Eq.(4), "Less-than" of Eq.(5) to compare for $tmp1 < tmp2$, and "Swap" of Eq.(5) to multiplex $tmp1$ and $tmp2$. In Step 2, we implement the modules for these nodes. We use FloPoCo tool [5] to generate an equation modules as a pipeline with floating-point operators. For the other modules, called HDL modules, we write HDL codes. In Step 3, we connect the modules considering their delays. To synchronize data flowing through the DFG, we have to adequately insert delay modules as shown in Fig.4. Although less delays are preferable in terms of resource consumption, it is not easy to find the best delay insertion when a lot of nodes are connected. In Step 4, we write an HDL code for the top module containing the stream processing core of the DFG and peripheral modules for Avalon-ST interface. Finally, in Step 5, we need to write a tcl-based configuration file containing the module information for the Qsys system integration tool.

The example of Fig.4 is quite simple, however, for example, fluid dynamics computation requires a complex DFG connecting much more nodes, resulting in a long development time. In addition to difficulty in finding adequate delay insertions, implementing and connecting a lot of modules are time-consuming and error-prone. To improve this poor productivity, we need a tool to automate Steps 2 to 5 and allow developers to concentrate on designing and exploring algorithms.

### IV. SPGEN: STREAM PROCESSOR GENERATOR

#### A. Overview of SPGen

We propose and develop a stream processor generator, called SPGen, to automate Steps 3 to 5 and a part of Step 2. Our requirements for SPGen are as follows:

1) Stream computing is given by its own simple and easy description with equations and module calls.
2) Users can write their own modules in HDL for non-numerical processing.
3) Users can control pipelining for design space exploration.
4) The output of SPGen is a core with Avalon-ST interfaces, which can be embedded into the SoC platform.

To easily describe stream computation for SPGen, we define a domain-specific stream processing description, called SPD. Although high-level descriptions like C language are not supported for now, we plan to develop code-to-code translation tools for them. In SPD, stream computation can be written with equations and module calls. SPGen generates equation modules by using the FloPoCo tool for equations. The combination of high-level equation description and low-level module-call description allows designers to easily implement a stream processor beyond just a floating-point (FP) pipeline.

SPGen is composed of a front-end and a back-end. In the following subsections, we describe the SPD format, the front-end, and the back-end.

#### B. Stream processing description (SPD) format

Fig.5 shows an SPD example of the stream processor for Eqs.(3) to (5). Each line follows a basic format of:

Label   item1, item2, item3;

The following labels are reserved for declaration or definition:

Name:   defines a top-module name (string).
Input:  declares input port variables (list with comma).
Output: declares output port variables (list with comma).
Param:  defines a constant FP value (paraleter = value).

If a label is not reserved word, the line declares a module by being interpreted as:

(Module name)   item1, item2, item3;

where

item1:  delay cycles of the module (integer).
item2:  module type, which is "equ" or "HDL" (string).
item3:  an equation with "=" or a module call (string).

Line 9 in Fig.5 declares an equation module "eq1" with variables $tmp1$, $a$, and $b$, and a parameter $p1 = 0.5$. An equation needs to be in a static single assignment form, where only a single variable can be written in the left-hand side. Since a pipeline delay is unknown before module generation, the delay cycles is ignored for an equation module.

Line 13 calls HDL module "less_than" with an instance name "lsthan." The delay cycle has to be given for an HDL module. Here, the "less_than" module has 1 cycle delay. An HDL module can have multiple output variables, which are specified with a variable list: (var1, var2, ...). The input variables are also specified with a variable list in parentheses.

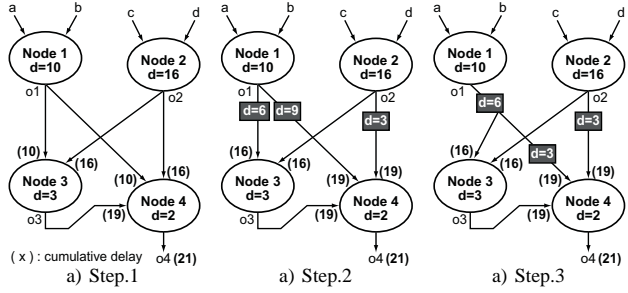

Fig. 6. Example steps of the delay insertion algorithm.

All variables in SPD description are basically 32-bit words, which are treated as IEEE754 single precision FP numbers for equation modules. We can use a bit range specification with [msb:lsb] for HDL modules. The used HDL module has to be prepared in advance. Developers can write their own modules while basic modules are available in a library.

### C. Front-end and back-end of SPGen

The front-end reads an SPD file and generates a DFG with nodes of modules as an intermediate expression. For the SPD description of Fig.5, such a DFG as Fig.3 is generated. The front-end parses lines one by one, and registers definitions and declarations in an intermediate data structure if a label is a reserved word. If not, a new node is created and set as an equation module or an HDL module with information of its input and output variables. After parsing all the lines, nodes with the same variable in an input variable list and an output variable list are connected with a directed edge. Finally, nodes having the same variables as the input and output ports of the top module are connected to the ports.

The back-end generates HDL files of equation modules, inserts delay modules into the DFG, and lastly outputs a Verilog-HDL code of the top module and a module configuration file. The back-end generates equation modules by using the FloPoCo tool [5] with equation descriptions. The tool generates pipelined computing modules with FP operators and outputs their HDL files, based on a target FPGA device and a target frequency passed by SPGen, which is used to determine the number of pipeline stages. Thus we can change the number of pipeline stages in a stream processor by specifying a target frequency via a command option of SPGen. Although the other parts than a stream processor may limit the actual operating frequency, we can explore design space under a trade-off between throughput and resource consumption. We can set a higher frequency for higher throughput with more stages while a lower frequency reduces resource consumption for pipeline registers and automatically-inserted delay modules.

After generating equation modules, their delay cycles are obtained. Then the back-end inserts delay modules into a DFG as shown in Fig.6, so that all the data inputs of each modules are synchronized. This delay insertion is performed based on the cumulative delays along paths from the top-module inputs. Because of the delay insertion algorithm, currently feed-back loops are not allowed in a DFG. Modules can have an initiation interval other than 1, however, developers are responsible for data synchronization in stream computation. After delay insertion, a Verilog-HDL file of the top module is generated. The back-end outputs codes for top-module declaration with input and output ports, module instances, and wires to connect the modules. Since the input and output ports of the top module include the signals of Avalon-ST interface, codes for these signals are also generated. Finally, the back-end generates a configuration file for the stream processor to be embedded into the SoC platform of Fig.1. The file contains tcl descriptions of module information, interface definitions, referenced HDL files, and so on.

## V. APPLICATION EXAMPLE AND EVALUATION

### A. HPC application example

For verification and evaluation of SPGen, we designed and implemented a stream processor for the lattice Boltzmann method (LBM) [8], [12], which is one of the methods to compute fluid dynamics. We compare it with manually implemented one in [12]. Here we summarize computations of the LBM stream processor while more details are available in [12]. LBM models fluids with propagation and collision processes of fictive particles over a discrete lattice mesh [4]. Each lattice cell has a distribution function $f_i$ for each of the nine particle speeds $c_i = \{(\cos \pi(i-1)/4, \sin \pi(i-1)/4) | i = 0, 1, ..., 8\}$. LBM consists of a macroscopic physical quantity calculation stage (Macro), an equilibrium calculation stage (Equi), a collision calculation stage (Col), a translation stage (Trans), and a boundary calculation stage (Bound). With these stages, $f_i(x)$ of a lattice cell at $x$ is updated for a single time step. Computations of these stages are given for $i = 0, ..., 8$ by:

$$\rho = \sum_{i=0}^{8} f_i(\boldsymbol{x}), \quad \boldsymbol{V} = \frac{1}{\rho} \sum_{i=0}^{8} \boldsymbol{c}_i \cdot f_i(\boldsymbol{x}), \quad (6)$$

$$f_i^{\text{eq}} = \rho \left\{ A_i + B_i(\boldsymbol{c}_i \cdot \boldsymbol{V}) + C_i(\boldsymbol{c}_i \cdot \boldsymbol{V})^2 - D_i \boldsymbol{V}^2 \right\} (7)$$

$$f_i^{\text{col}}(\boldsymbol{x}) = f_i(\boldsymbol{x}) - \frac{f_i(\boldsymbol{x}) - f_i^{\text{eq}}(\boldsymbol{x})}{\tau}, \quad (8)$$

$$f_i^{\text{tr}}(\boldsymbol{x} + \boldsymbol{c}_i) = f_i^{\text{col}}(\boldsymbol{x}), \quad (9)$$

where $\rho$, $\boldsymbol{V} = (u, v)$, and $\tau$ are fluid density, fluid velocity, and a single relaxation time. $A_i$, $B_i$, $C_i$ and $D_i$ are constants for $i = 0, 1, ..., 8$. Macro stage computes Eqs.(6). Equi, Col, and Trans stages computes Eqs.(7), (8), (9), respectively, for $i = 0, 1, ..., 8$. Then, boundary conditions are applied to $f_i^{\text{tr}}$ for boundary cells, which are defined with attribute flags of each cell. We can compute boundary conditions for solid surfaces or fluid-incoming/outgoing with constant pressures. Details of boundary conditions are described in [12].

### B. Implementation, validation, and productivity improvement

We wrote an SPD code for the LBM stream processor to compute Eqs. (6) to (9) and the boundary conditions. Fig.7 is an excerpt of the SPD code. Input and Output define the input and the output of the stream computation, respectively. The nine inputs of "if0" to "if8" are numerical values in IEEE754 floating-point format. The input of "iAtr_RAW_" is an attribute of each cell, which contains flags for boundary processing. Here an Input/Output port name with "_RAW_" is treated as non-numerical 32-bit words, and therefore an internal floating-point format converter between IEEE754 (32 bits) and FloPoCo (34 bits) is not inserted. The ports names with "_VLD_", "_SOP_", and "_EOP_" are internal signals of valid, start of packet, and end of packet for Avalon-ST interface. If HDL modules in SPD description use them, the above ports have to be explicitly written, and connected to the modules. If not, they can be omitted while their input signals are automatically connected to the output ports by SPGen.

In this design, we used HDL modules of a delay module "mDelay", a multiplexer module "mMux", and a translation

```
Name     mLBM_allStages;
Input    if0,if1,if2,if3,if4,if5,if6,if7,if8,iAtr_RAW_,i_VLD_,i_SOP_,i_EOP_;
Output   of0,of1,of2,of3,of4,of5,of6,of7,of8,oAtr_RAW_,o_VLD_,o_SOP_,o_EOP_;

Param    P_rho_in         = 1.05;
Param    P_rho_out        = 0.95;
Param    P_one_tau        = 0.516262261;
Param    P0               = 0.666666666666667;
# ...  (Similar lines skipped for Param P1,P2,P3,P4,P5,P6,P7,P8,P9,Pa, and Pb)

uoAtr    1,HDL, oAtr_RAW_ = mDelay(Atr_tr),<.pWidth(34),.pDelay(1)>;

#-------- Macro calc stage ----------------------------------
if5Mif7  0,equ,   if5Mif7  = (if5 - if7);
if6Mif8  0,equ,   if6Mif8  = (if6 - if8);
if1Mif3  0,equ,   if1Mif3  = (if1 - if3);
if2Mif4  0,equ,   if2Mif4  = (if2 - if4);
rho      0,equ,   rho      = (((if5+if7)+(if6+if8)) + ((if1+if3)+(if2+if4)+if0));
rho_u    0,equ,   rho_u    = ( (if5Mif7-if6Mif8) + if1Mif3);
rho_v    0,equ,   rho_v    = ( (if5Mif7+if6Mif8) + if2Mif4);
rho_uMv  0,equ,   rho_uMv  = (  if1Mif3-if2Mif4) - (2.0*if6Mif8);
rho_uPv  0,equ,   rho_uPv  = (  if1Mif3+if2Mif4) + (2.0*if5Mif7);

#-------- Equlibrium calc stage -----------------------------
rho_u2   0,equ,   rho_u2   = (rho_u   * rho_u  );
rho_v2   0,equ,   rho_v2   = (rho_v   * rho_v  );
rho_uPv2 0,equ,   rho_uPv2 = (rho_uPv * rho_uPv);
rho_uMv2 0,equ,   rho_uMv2 = (rho_uMv * rho_uMv);
rho2     0,equ,   rho2     = (rho_u2  + rho_v2 );
divrho   0,equ,   divrho   = (1.0 / rho);
f0eq     0,equ,   f0eq     = ((Pa*rho)-             (Pb*rho2) *divrho);
f1eq     0,equ,   f1eq     = ((P6*rho)+(P3*rho_u))+(((P2*rho_u2)-(P4*rho2))*divrho);
# ...  ( Similar lines skipped for f2eq, f3eq, f4eq, f5eq, f6eq, f7eq, f8eq )

#-------- Collision Stage -----------------------------------
f0_co    0,equ,   f0_co    = if0 - P_one_tau * (if0 - f0eq);
# ...  ( Similar lines skipped for f1_co to f8_co )

#-------- Translation stage ---------------------------------
uTrWrap 1506,HDL,(f0_cp,f1_tr,f2_co,f3_tr,f4_cp,f5_tr,f6_tr,f7_tr,f8_tr,Atr_tr, \\
                                                    o_VLD_,o_SOP_,o_EOP_) \\
     = mTrans(f0_co,f1_co,f2_co,f3_co,f4_co,f5_co,f6_co,f7_co,f8_co,iAtr_RAW_, \\
                                 i_VLD_[0],i_SOP_[0],i_EOP_[0]),\\
                         <.pWordWidth(34),.pUnitLength(360),.pSelLen(3'b011)>;

#-------- Const-press stage ---------------------------------
uf6_tmp  1,HDL,   f6_tmp   = mMux(f6_tr, f8_tr, Atr_tr[11]), <.pWidth(34)>;
uf3_tmp  1,HDL,   f3_tmp   = mMux(f3_tr, f1_tr, Atr_tr[11]), <.pWidth(34)>;
uf7_tmp  1,HDL,   f7_tmp   = mMux(f7_tr, f5_tr, Atr_tr[11]), <.pWidth(34)>;
urho_in  0,equ,   rho_in   = ( P_rho_in  );
urho_out 0,equ,   rho_out  = ( P_rho_out );
urhoGiven 1,HDL, rhoGiven = mMux(rho_in, rho_out, Atr_tr[11]), <.pWidth(34)>;
urho_diff 0,equ, rho_diff = ( rhoGiven - ((f0_cp+f2_co) + (f3_tmp+f4_cp) \\
                                                       + (f6_tmp+f7_tmp)) );
uf1_cp_tmp 0,equ, f1_cp_tmp = ( P0 * rho_diff );
uf5_cp_tmp 0,equ, f5_cp_tmp = ( P1 * rho_diff );
uf1_cp   1,HDL,   f1_cp    = mMux(f1_tr, f1_cp_tmp, Atr_tr[9]), <.pWidth(34)>;
# ...  ( Similar lines skipped for uf5_cp, uf8_cp, uf3_cp, uf7_cp, and uf6_cp )

#-------- Bounce-back stage ---------------------------------
uof0     0,equ,   of0 = f0_cp;
uof1     1,HDL,   of1 = mMux( f1_cp, f3_cp, Atr_tr[3] ), <.pWidth(34)>;
# ...  ( Similar lines skipped for uof2 to uof8 )
```

Fig. 7. SPD code of the stream processor for LBM (excerpt of original 83 lines without empty or comment lines).

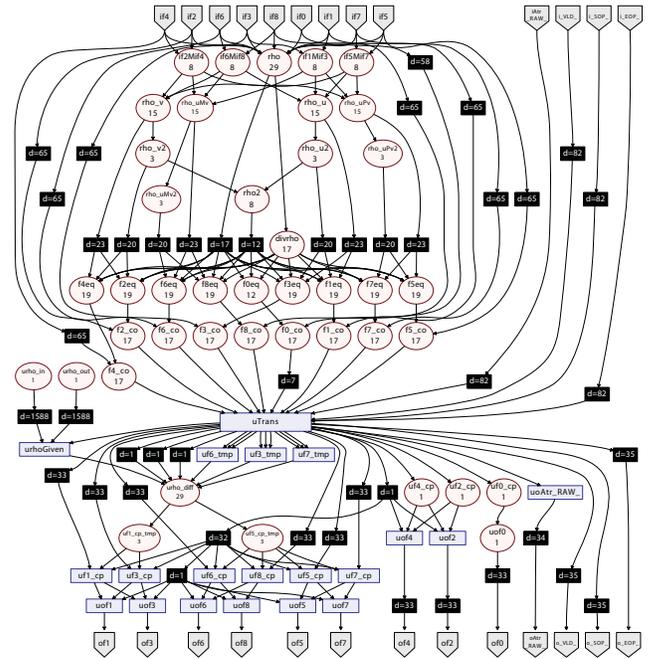

Fig. 8. SPGen compiled results: DFG of the stream processor for LBM.

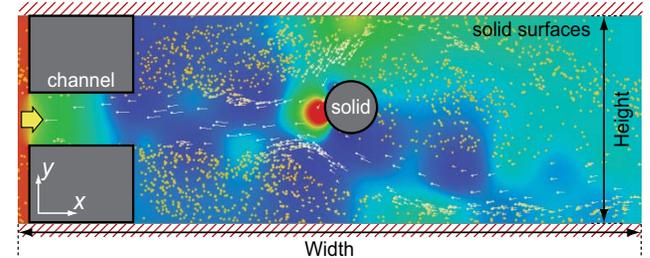

Fig. 9. Computed 2D sudden-expansion channel flow with a solid circle.

module "mTrans." In the case of HDL module declaration, a parameter list for a Verilog module instance can be added such as written for "mDelay" module. For example, we can change the lattice width to be computed by the stream processor for LBM by setting the parameters to the mTrans module. The SPD file for the LBM stream processor has only 83 lines without any empty or comment lines.

We compiled the SPD code with SPGen. Fig.8 shows the DFG of the generated processor. The circles are the equation modules while the rectangulars are the HDL modules. The filled rectangulars are delay modules automatically inserted by SPGen. We embedded the stream processor into the FPGA-based SoC platform of Fig.1, generated the system by using Qsys, and compiled the entire design for Stratix IV EP4SGX230 FPGA [1] on TERASIC DE4-230 board [3] with ALTERA Quartus II ver 13.1. We used "speed" option for place and route. The stream processor for LBM worked completely. Fig.9 shows the computational result obtained with the stream processor running on the FPGA.

To evaluate productivity improvement, we compare the development period and the number of lines in codes without any comment and empty lines between hand-designed implementation [12] and the implementation in SPD. In the case of the hand-designed LBM processing element (PE), even only the top modules of the LBM stages required seven files, which totally contain 2720 lines. This means that the SPD format reduced the lines to only 83 lines for SPGen. For development of the hand-designed PE, we spent several months for the conventional design flow. On the other hand, it took only one or a few days to write and debug the SPD code, and run the design on FPGA. Although we cannot say that this is complete comparison for quantitative evaluation of productivity, these results show some of productivity improvement by SPGen. Consequently, we could optimize the equations of LBM by removing some redundant operations in the SPD code.

The LBM stream processor generated by SPGen has the same computing performance as that of the hand-designed one. Since the pipelining overhead for epilogue and prologue is ignorable for sufficiently large lattice data, we can obtain the sustained performance by multiplying an operating frequency $F$ with the number of FP operations per cycle $N_{\text{ops}}$. In our system, $F = 125$ MHz and $N_{\text{ops}} = 131$ for 70 adders, 60 multipliers, and 1 divider in the stream processor. As a result, the sustained performance is $FN_{\text{ops}} = 16.4$ GFlops. Since at most three stream processors fit the Stratix IV FPGA, expected maximum performance is $16.4 \times 3 = 49.2$ GFlops.

TABLE I. RESOURCE CONSUMPTION FOR TARGET FREQUENCIES OF 125, 250, AND 500MHZ, AND COMPARISON WITH THE DESIGN BY HAND [12].

| Target frequency | Module | Combinational LUTs | % | Registers | % | Block memory Kbits | % | 36-bit DSPs | % |
|---|---|---|---|---|---|---|---|---|---|
| Stratix IV EP4SGX230 | | 182400 | 100 | 182400 | 100 | 14283 | 100 | 322 | 100 |
| 125 MHz | Peripherals | 42758 | 23.4 | 43491 | 23.8 | 2736 | 19.2 | 0 | 0 |
| 125 MHz | Stream processor | 44693 | 24.5 | 20745 | 11.4 | 1204 | 8.4 | 15 | 4.7 |
| 250 MHz | Peripherals | 42868 | 23.5 | 43525 | 23.9 | 2736 | 19.2 | 0 | 0 |
| 250 MHz | Stream processor | 35411 | 19.4 | 33533 | 18.4 | 1233 | 8.6 | 15 | 4.7 |
| 500 MHz | Peripherals | 42931 | 23.5 | 42692 | 23.4 | 2740 | 19.2 | 0 | 0 |
| 500 MHz | Stream processor | 45284 | 24.8 | 56001 | 30.7 | 1319 | 9.2 | 15 | 4.7 |
| Design by hand [12] | Peripherals | 37942 | 20.8 | 41316 | 22.7 | 2748 | 19.2 | 0 | 0 |
| Design by hand [12] | LBM PE | 46889 | 25.7 | 38117 | 20.9 | 1272 | 8.9 | 74 | 23.0 |

*C. Resource consumption and design space exploration*

By changing the number of pipeline stages of a stream processor with different target frequencies for the FloPoCo tool, we explore the design space to find one for minimum resource consumption. Table I shows the resource consumption for target frequencies of 125MHz, 250MHz, and 500MHz. "Peripherals" are for PCI-Express core, two DDR2 memory controllers, and four DMA cores. Here no microprocessor is implemented. The table shows that the higher frequency requires more registers for increased pipeline stages. Simultaneously, the higher frequency also consumes more block memory bits because more delays are inserted for a more deeply pipelined DFG. On the other hand, the target frequency of 250MHz gives the smallest number of LUTs against our expectations. This is because the FloPoCo tool uses smaller FP operators for the 250 MHz design. In any case, SPGen allows us to easily find the best design from pipelining choices.

We also compares resource consumption with the hand-designed PE, which is functionally identical. As seen in Table I, The 250MHz design has smaller consumption especially for LUTs and DSPs. Less LUTs are achieved mainly by the equation optimization where redundant operators are removed in the SPD code. Less DSP blocks are due to constant multipliers. In the hand-designed PE, we store parameters in registers to be read and computed. Therefore normal multipliers are generated for them. On the other hand, we use constant parameters directly in the SPD code, and change them when necessary. This is because SPGen allows us quick modification and synthesis of the design. As a result, many multipliers are generated as constant ones with smaller logics.

## VI. CONCLUSIONS

This paper presents SPGen, a stream processor generator, which is based on the stream computing model and aims at assisting developers to design and implement high-throughput reconfigurable stream processor on an FPGA-based common SoC platform. We define a domain-specific description format, SPD, to easily define stream computation with equations and module calls. SPGen allows us to focus on designing stream algorithms and explore design space for different equations and frequencies. We demonstrate and validate SPGen and the SPD format with fluid dynamics computation based on LBM. SPGen reduces a development time from months to a few days. The generated stream processor works completely, and its performance and resource consumption are comparable to or better than those of the hand-designed one.

In our future work, we will evaluate SPGen with other applications of high-performance or embedded computation. We will extend SPGen to version 2.0 so that it can generate a stream processor with interfaces for inter-processor communication other than data streaming.

ACKNOWLEDGMENTS

This research was partially supported by Grant-in-Aid for Scientific Researches (B) No.23300012 and No.25280041, and for Challenging Exploratory Research No.23650021 from the Ministry of Education, Culture, Sports, Science and Technology, Japan. We thank support by ALTERA university program.